\documentclass[prd,showkeys,preprintnumbers,amsmath,amssymb,floatfix]{revtex4}

\usepackage[utf8]{inputenc}
\usepackage{hyperref}

\usepackage{dcolumn}
\usepackage{bm}
\usepackage{graphicx}
\usepackage{subfigure}
\usepackage[usenames,dvipsnames,svgnames]{xcolor}

\def\be{\begin{equation}}
\def\ee{\end{equation}}
\def\bestar{\begin{equation*}}
\def\eestar{\end{equation*}}

\begin{document}

\title{Particle production from noncommutative-inspired Reissner-Nordstr\"om black hole}

\author{Alberto Maya}
\email{josealbertom66@nu.ac.th}
\affiliation{The Institute for Fundamental Study, Naresuan University,\\ 
Phitsanulok 65000, Thailand}

\author{Marco Maceda}
\email{mmac@xanum.uam.mx}
\affiliation{Departamento de F\'{i}sica \\ Universidad Aut\'{o}noma Metropolitana - Iztapalapa\\
                   Av. Ferrocarril San Rafael Atlixco 186, C.P. 09310, México City, M\'exico}                   
                                 
\date{\today}

\begin{abstract}
By considering a massless Klein-Gordon field and the moving mirror model, we calculate particle production for a noncommutative-inspired Reissner-Nordström black hole sourced by smeared distributions of mass and charge. We show that noncommutative truncated trajectories diminish the emission spectrum at large frequencies, due to noncommutative effects on the external surface gravity of the black hole, in sharp contrast with the spectrum of a classical Reissner-Nordström black hole.

\end{abstract}


\maketitle

\section{Introduction}
\label{intro}

Soon after it was realised that a black hole possesses a temperature and that a black hole entropy could be defined in terms of the horizon area, the laws of thermodynamics of black holes were formulated~\cite{Bardeen:1973gs}; it was expected then that thermodynamical quantities, such as entropy, could have a sound meaning within the context of black holes. 

As it is well known, the temperature of a black hole is related to the surface gravity associated with its horizon. This temperature can be visualised as a flux of particles that an observer at spatial infinity may detect as radiation; in the situation where there is more than one horizon, radiation is usually associated with physical processes happening at the outer horizon. The discovery by Hawking~\cite{Hawking:1975vcx} that a black hole could evaporate by emitting radiation represented a milestone that led to new questions about its nature. On one hand, it motivated a deeper discussion of the flux of particles measured by different observers and how the notion of a vacuum state in field theory differs drastically from one observer to another in curved spacetime~\cite{Davies:1974th,Wald:1975kc,Unruh:1976db}. At the same time, it also raised questions about information loss during the evaporation process, an issue that continues to be explored nowadays~\cite{Page:1993wv,Ryu:2006bv,Ryu:2006ef,Page:2013dx,Almheiri:2019qdq,Almheiri:2019hni,Penington:2019kki,Maldacena:2020ady}.

As a quantum theory of gravity is still in the works, it is vital to consider frameworks in which quantum effects may be incorporated into gravity with the aim of understanding its relevance in physical processes. Noncommutative geometry is one of these approaches, which finds a natural application in General Relativity when addressing the issue of singularities in black hole solutions. By implementing commutation relations among coordinates, we can introduce quantum effects such that a classical point-like behaviour is now replaced by a smeared distribution; in this fashion, quantities like the curvature of a black hole become regularised because of a de Sitter core in the region close to the location of the black hole~\cite{Spallucci:2014kua,Nicolini:2005vd,Nicolini:2005gy,Banerjee:2009xx}. 

In this work, we discuss the phenomenological implications for particle production associated with a black hole when a noncommutative structure in spacetime exists. For this purpose, we consider a noncommutative-inspired Reissner-Nordström (RN) black hole sourced by smeared noncommutative distributions of mass and charge~\cite{Nicolini:2008aj}. Since noncommutativity modifies the location of the event horizon of a black hole and the strength of the associated surface gravity, we expect that it has a direct influence on the radiation emitted by it. In other complementary approaches to this issue, the noncommutative features of a black hole solution can also be addressed from a thermodynamic perspective. Specifically, we can quantify noncommutative effects through the noncommutative parameter, which also obeys thermodynamic laws and has been considered as an extensive thermodynamic variable playing the role of an effective charge; such an interpretation provides insight into its thermodynamic behaviour in Hawking radiation \cite{Nozari:2006rt, Maya:2026gaz}. 

We organise this paper as follows: in Section~\ref{secc:2}, we briefly review the construction of the noncommutative-inspired RN black hole sourced by smeared distributions of matter and charge. The key aspects of the moving mirror model in this modified gravitational background are presented in Section~\ref{secc:3}, highlighting the difference with classical trajectories. Afterwards, the Bogoliubov coefficient associated with particle emission is calculated in Section~\ref{secc:4}; there, we also determine the resulting particle spectrum and compare it with the classical one. We end with some remarks and perspectives in the Conclusions section.

\section{Noncommutative-inspired Reissner-Nordstr\"om black hole}
\label{secc:2}

In this section, we briefly review the construction of noncommutative-inspired black holes based on noncommutative distributions of mass and charge. To begin with, we consider in a $(d+1)$-dimensional spacetime a set of commutative relations among spatial position operators 
\be
[x^i, x^j] = i \theta^{ij}, \quad i, j = 1, \dots, d,
\ee
with elements $\theta^{ij}$ that are constant and antisymmetric, i.e. $\theta^{ij} = -\theta^{ji}$. Of particular interest is the case where all the elements $\theta^{ij}$ with $i < j$ are equal to a common value $\theta$ because this choice leads to a partition of the spacetime in 2-dimensional hyperplanes with basically the same noncommutative structure; it can be shown that in each of these hyperplanes, a class of coherent states that are complete with respect to the so-called Voros product exists~\cite{Banerjee:2009xx}. Furthermore, the coherent states defined on them are the natural states to be used when calculating average values for observables; in particular, the average values of the position operators associated to one of these hyperplanes. It may be shown that these expectation values are given by smeared distributions  
\be
\rho (\bm r) = \frac 1{(4\pi\theta)^{d/2}} e^{-r^2/4\theta}.
\ee
These smeared distributions replace the classical description of points in spacetime; consequently, particles in this background are no longer localized, and we now speak of a fuzzy spacetime. It is straightforward to see that in the limit $\theta \to 0$, we recover from the smeared distributions the standard commutative result $\rho(\bm r) \sim \delta^{(d)}(\bm r)$ associated with a point-like description.

The previous modification, from a point-like description to an extended structure, was first noted in the context of noncommutative quantum field theory~\cite{Smailagic:2003rp,Smailagic:2004yy}; later, it was suggested as a way to regularise essential singularities in classical solutions to Einstein's field equations describing black holes. Along this direction, the first application was the noncommutative Schwarschild solution~\cite{Nicolini:2005vd}, followed by the noncommutative Reissner-Nordstr\"om solution~\cite{Nicolini:2008aj}. Later, by exploiting the Kerr-Schild structure of the Kerr and Kerr-Newman solutions, the procedure was generalised to obtain noncommutative inspired rotating black holes~\cite{Modesto:2010rv}. To illustrate this procedure, we consider in the following the specific case of the noncommutative inspired Schwarzschild black hole. The starting point is a spherically symmetric metric
\be
ds^2 = -f(r)dt^2 + \frac1{f(r)}dr^2 + r^2 (d\theta^2 + \sin^2 \theta d\phi^2).
\ee 
from which the left-hand side of the noncommutative gravitational field equations
\be
R_{\mu\nu} - \frac 12 g_{\mu\nu} R = 8\pi\, {\mathfrak T}_{\mu\nu}
\label{ncfieldequs}
\ee
is calculated; here the Ricci tensor $R_{\mu\nu}$ and the scalar curvature $R$ have their usual meanings. On the other hand, the noncommutative energy-momentum tensor
\be
{\mathfrak T}^\mu_\nu \equiv \mbox{diag} (-\rho(r), p_r(r), p_t(r), p_t(r)), \qquad p_r (r) \equiv -\rho (r) \equiv -\frac m{(4\pi \theta)^{3/2}} e^{-r^2/4\theta}, \qquad p_t (r) \equiv -\rho(r) - \frac r2 \rho(r)_{,r}. 
\ee
provides the right-hand side of the field equations; it is written in terms of the noncommutative smeared distribution $\rho(r)$ and represents an anisotropic fluid \cite{Nicolini:2008aj}. It is worth mentioning that according to this approach, noncommutativity affects only the energy-momentum tensor whilst geometrical objects remain unchanged. Notice that $\int_V  \rho(r) d^3 r = m$ when we integrate over the whole space; clearly in the limit $\theta \to 0$, we recover the point-like expression $\rho(r) = m\,\delta^{(3)}({\bm r})$, which allows us to identify $m$ with the mass of the black hole. This identification of the parameter $m$ gets reinforced by solving the field equations; 
the $tt$-component provides the relevant field equation 
\be
(r f)_{,r} = 1 - \frac m{\sqrt{\pi \theta^3}} e^{-r^2/4\theta}r^2,
\ee 
with solution
\be
f = 1 - \frac {4m}{\sqrt{\pi} r} \gamma \left( \frac 32, \frac {r^2}{4\theta} \right).
\ee
Here $\gamma(a, z)$ is the lower incomplete gamma function; it is easy to see that $f = 1 -2m/r$ in the limit $\theta \to 0$.

The extension of this procedure to obtain a noncommutative-inspired charged solution for Maxwell electrodynamics coupled to the gravitational field is straightforward: first, we define a noncommutative-inspired 4-current
\be
{\cal J}^\mu \equiv \rho_e (r)  \,\delta^\mu_0 \equiv \frac q{(4\pi \theta)^{3/2}} e^{-r^2/4\theta} \,\delta^\mu_0,
\ee
where $q$ represents the charge of the black hole. Then, by solving the noncommutative conservation laws
\be
F^{\mu\nu}{}_{; \mu} = 4\pi {\cal J}^\mu, 
\ee
where $F_{\mu\nu} \equiv \partial_\mu A_\nu - \partial_\nu A_\mu $ is the standard electromagnetic tensor, we obtain the associated noncommutative 4-potential $A_\mu$, which in turn is used to calculate the components of the noncommutative-inspired energy-momentum tensor
\be
T_{e.m.}^{\mu\nu} \equiv -2 F^{\mu\sigma} F^\nu{}_\sigma + \frac 12 g^{\mu\nu} F^{\alpha\beta} F_{\alpha\beta}.
\ee
Adding $T_{e.m.}^{\mu\nu} $ to the right-hand side of the gravitational field equations in Eq.~(\ref{ncfieldequs}) leads to~\cite{Nicolini:2008aj}
\be
f = 1 - \frac {4m}{\sqrt{\pi} r} \gamma \left( \frac 32, \frac {r^2}{4\theta} \right) + \frac {q^2}{\pi r^2} \left[ \gamma^2 \left( \frac 12, \frac {r^2}{4\theta} \right) - \frac r{\sqrt{2\theta}}\gamma \left( \frac 12, \frac {r^2}{2\theta} \right) + \sqrt{\frac 2\theta}r \gamma \left( \frac 32, \frac {r^2}{4\theta} \right) \right].
\label{metricf}
\ee 
From this expression we recover the classical RN metric in the commutative limit $\theta \to 0$; therefore, the above solution is known as the noncommutative-inspired RN black hole~\cite{Nicolini:2008aj}. In the next sections, we discuss particle production in this gravitational background. 

\section{Mirror model and truncated trajectories}
\label{secc:3}

Particle production associated with black hole radiation can be understood as a quantum tunnelling effect~\cite{Parikh:1999mf,Chakraborty:2014gwx} or in terms of the so-called moving mirror (MM) model in flat spacetime~\cite{Moore:1970tmc,DeWitt:1975ys,Davies:1976hi,Davies:1977yv,Carlitz:1986nh,Wilczek:1993jn,Good:2013lca,Good:2015jwa}. In the latter, Hawking radiation can be analysed in Minkowski spacetime when an MM is introduced as a dynamical surface with appropriate boundary conditions~\cite{Davies:1976hi,Davies:1977yv}. The MM model in Minkowski spacetime imitates the whole behaviour of black hole radiation, so that it is possible to study particle emission through it.

To be more precise, in curved spacetime, we can consider a quantum field satisfying a regularity condition at the origin; these field modes are then put into one-to-one correspondence with those associated with a moving mirror in (1+1)-Minkowski spacetime. This one-to-one correspondence establishes a link between particle production in a black hole and that due to an MM in flat spacetime. In the following, we work along the lines of~\cite{Good:2020qsy}, where the MM formalism is employed to discuss particle production in the classical RN spacetime. As we proceed, the appropriate changes for the noncommutative-inspired RN solution will be indicated.

\begin{figure}[ht]
\includegraphics[width=0.6\textwidth]{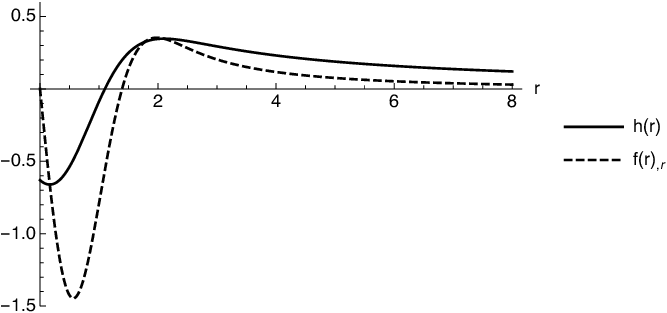}
\caption{Plot of $h (r)$ versus $r$ (continuous line) and $f(r)_{,r}$ versus $r$ (dashed line) for a noncommutative-inspired RN black hole with parameters $m=1, q=0.5$ and $\theta = 0.2$. Both curves coincide at the external horizon $r_+ = 1.772$, so that $h_+ = \lim_{r\to r_+} h(r)$ gives the value of the external surface gravity.}
\centering
\label{funch}
\end{figure}

As mentioned before, let us consider a massless scalar field $\phi_\omega$ in (1+1)-Minkowski spacetime which satisfies the Klein-Gordon equation $\Box \phi_\omega= 0$ and the Dirichlet condition $\phi_\omega (t, z(t))=0$, where $z=z(t)$ is the MM trajectory function. This equation defines a reflecting condition since, at the mirror location, the scalar field vanishes. Within the MM model, field modes are then excited, leading to particle production, and therefore, given a suitable trajectory, it is possible to replicate the particle production associated with a collapsing star. In fact, for a rapidly receding mirror, incoming rays become strongly red-shifted, similar to light rays escaping from the black hole horizon~\cite{Wilczek:1993jn}. A detailed analysis shows that the Bogolyubov coefficient $\beta_{\omega \omega^\prime}$ linked to particle emission admits the following integral representation~\cite{Good:2016atu,Good:2013lca}
\begin{equation}
\beta_{\omega \omega^\prime}= \frac{1}{4 \pi \sqrt{\omega \omega^\prime}} 
\int_{-\infty}^{v_0} dv e^{-i \omega^\prime v -i \omega z(v)} \left(\omega^\prime - \omega \frac{dz(v)}{dv}\right).
\label{bogocoeff}
\end{equation}
Here $z(v)$ denotes the MM trajectory as a function of the advanced null coordinate $v \equiv t + r^*$. Let us at this point recall that in the commutative scenario~\cite{Good:2020qsy}, the Bogolyubov coefficient $\beta_{\omega \omega^\prime}$ in Eq.~(\ref{bogocoeff}) may be evaluated using a tortoise coordinate $r^*$ defined by $dr^*/dr \equiv f^{-1}$.
Unlike the classical case, writing a tortoise coordinate $r^*$ for a noncommutative spherically symmetric metric is not straightforward as before. Since the tortoise coordinate requires the integration of $f^{-1}$ with respect to the radial coordinate, and recalling that the noncommutative RN metric function $f$ in Eq.~(\ref{metricf}) contains incomplete gamma functions, the calculation of the noncommutative tortoise coordinate poses some technical issues. To deal with them, it is convenient to rewrite the metric function in such a way that we can isolate its singular and nonsingular behaviour at the horizon location $r=r_+$, where $f(r_+) = 0$. For this purpose, we first define the following auxiliary function
\begin{equation}
h(r) \equiv  \frac{f(r)}{r-r_+}. 
\end{equation}
This function $h$ has a well-defined value at $r=r_+$; in fact, if we define $h_+ \equiv \lim_{r \to r_+}h(r)$, it follows that $ h_+ = f_{,r} (r_+)$; therefore, $h_+$ is proportional to the surface gravity $\kappa_+$ at the external horizon. In Fig.~\ref{funch} we show the behaviour of $h$ for a typical noncommutative RN black hole.

Using now this function $h$, together with its Taylor expansion around  $r=r_+$, we can write
\begin{equation}
\frac 1f = 1 + \frac{1}{h_+(r - r_+)} + \frac{1}{h_+(r - r_+)}\left[\sum_{k=1}^\infty a_k (r-r_+)^k - h_+(r - r_+)\right], 
\label{invf}
\end{equation}
where the coefficients $a_k$ are related to those appearing in the Taylor series expansion for $h$ at $r=r_+$. Once we separate the metric function according to Eq.~(\ref{invf}), we can calculate the tortoise coordinate $r^*$ from its definition; thus, we find
\begin{equation}
r^* = r + \frac{1}{2 \kappa_+} \ln \left|\frac{r-r_+}{r_+}\right| + A(r), 
\end{equation}
where 
\begin{equation}
A(r) \equiv \frac{1}{h_+} \int^r \frac{ds}{s-r_+} \left[ \sum_{k=1}^\infty a_k (s-r_+)^k - h_+(s - r_+) \right],
\end{equation}
is an analytic function at $r = r_+$. We see from this expression that the structure of the tortoise coordinate of the noncommutative-inspired RN black hole is similar to the commutative one; the only meaningful difference is given by the value of the surface gravity $\kappa_+$, where noncommutative contributions come into play.

\begin{figure}[ht]
\includegraphics[width=0.45\textwidth]{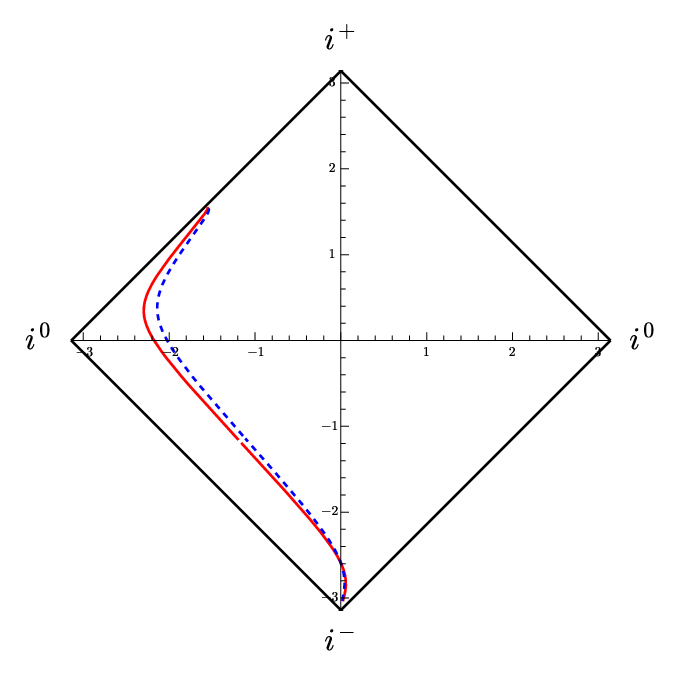}
\caption{Penrose diagram showing generic truncated MM trajectories in the noncommutative-inspired RN black hole (continuous red line) and the classical RN black hole (dashed blue line) with $m=1, q=0.5$ and $\theta=0.2$. Trajectories coincide in the far past and future, with a noticeable difference at intermediate times due to the value of the external surface gravity in each model.}
\centering
\label{mmpath}
\end{figure}

Once our noncommutative tortoise coordinate has been found, we introduce the matching condition~\cite{Wilczek:1993jn} between external and internal null coordinates through the relation
\begin{equation}
r^* \left(r=\frac{v_0 - U}{2}\right)= \frac{v_0 - u}{2}, 
\end{equation}
where $v_0$ can be chosen conveniently as $2r_+$ if $v_+ = 0$. Then, the function $u(U)$ for the internal region can be used to define our mirror trajectory $z(v)$ like in the classical solution through the identifications $U \leftrightarrow v, u(U) \leftrightarrow z(v)$ ~\cite{Good:2020qsy}; we obtain thus
\begin{equation}     
z(v)= z_{Trun} + B(v), 
\label{mmtrajectory}
\end{equation}
where $z_{Trun} \equiv v - \frac{1}{\kappa_+} \ln \left(\frac{1}{2} \sqrt{\frac{v^2}{r_+^2}}\right)$, and $B(v) \equiv - 2A\left(r=r_+ - \frac{v}{2}\right)$. It is worth emphasising that the function $B(v)$ inherits properties of $A(r)$; in particular, it does not have any singularities. On the other hand, the function $z_{Trun}$ has a singularity at $v=0$; furthermore, it represents the main contribution to the particle emission by the black hole, since for intermediate times, it fully reproduces the behaviour of the MM trajectory $z(v)$ with only slight deviations at later times that become negligible as $t \rightarrow +\infty$. The truncated trajectory is defined here as a path that neglects the contribution related to the inner horizon. Thus, the meaningful behaviour of an MM trajectory can be given even if the contributions of the $B(v)$ function are removed. Fig.~\ref{mmpath} shows a generic truncated MM trajectory in the noncommutative-inspired RN black hole contrasted with the classical one. Even though the paths are close in shape, there is a difference arising from the different values of the surface gravity at the external horizon.

\section{Bogoliubov $\beta_{\omega\omega'}$ coefficient and particle spectrum}
\label{secc:4}

\begin{figure}[ht]
\includegraphics[width=0.5\textwidth]{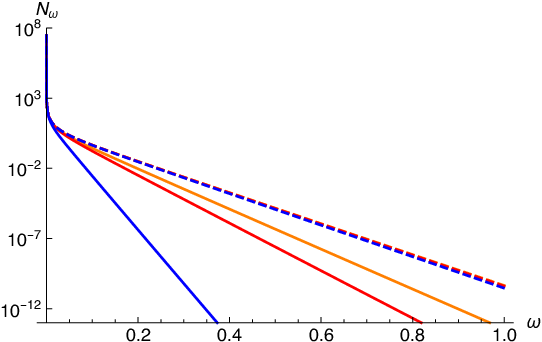}
\caption{Plot of the particle spectrum for the noncommutative-inspired RN black hole with $m=1, \theta=0.2$ for $q=0.3$ (orange line), $q=0.5$ (red line) and $q=0.65$ (blue line). The dashed lines, practically indistinguishable from each other, give the particle spectrum of the corresponding classical RN black holes for the same set of values of the charge. Notice that particle production is lower in the noncommutative case compared with the classical situation and strongly suppressed for large frequencies.} 
\centering
\label{partprod}
\end{figure}

From Eq.~(\ref{bogocoeff}), we can determine the Bogolyubov coefficient $\beta_{\omega \omega^\prime}$ as long as the whole MM trajectory $z(v)$ is specified. For the particular case of the noncommutative RN spacetime, we use Eq.~(\ref{invf}) in Eq.~(\ref{bogocoeff}) and integrate by parts, neglecting boundary terms and choosing $v_+=0$. Additionally, we must consider that the main contributions are governed by the truncated MM trajectory \cite{Good:2020qsy}, since the remaining contributions encoded in the function $B(v)$ in Eq.~(\ref{mmtrajectory}), rapidly oscillate as $v$ increases; consequently, the main contribution comes from the region in which $v$ takes values close to zero. Thus, with this in mind, we obtain the following form for the Bogoliubov coefficient 
\begin{equation}
\beta_{\omega \omega^\prime}^{Trun} \approx \frac{\omega^\prime}{2 \pi \sqrt{\omega \omega^\prime}} \int_{-\infty}^{0} e^{-i\omega^\prime v} e^{-i\omega z_{Trun}} dv.  
\end{equation}
This result extends the previously known expression in~\cite{Good:2020qsy} for a commutative spherically symmetric metric; it also helps us to understand why the relevant physics is captured by the function $z_{Trun}$ in the noncommutative scenario.
Calculating the Bogoliubov coefficient explicitly, we encounter
\begin{equation}
\beta^{Trun}_{\omega \omega^\prime}=\frac{1}{2\pi} \sqrt{\frac{\omega^\prime}{\omega}} \left(\frac{1}{2r_+}\right)^{i \omega/\kappa_+} \frac{e^{\frac{i\pi}{\kappa_+}}}{\left[ -i(\omega + \omega^\prime)\right]^{1+i\omega/\kappa_+}} \Gamma \left( 1+ \frac{i \omega}{\kappa_+} \right).
\end{equation}
Once we have determined the Bogoliubov coefficient $\beta_{\omega \omega^\prime}$, we can find the number of emitted particles 
~\cite{Good:2013lca}
\begin{equation}
N_\omega=\int_{0}^{\infty} |\beta_{\omega \omega^\prime}|^2 d \omega^\prime, 
\end{equation}
by performing a similar calculation as in the commutative case; we obtain thus
\begin{equation}
N_\omega = \frac{1}{2 \pi \kappa_+} \frac{1}{e^{\frac{2 \pi\omega}{\kappa_+}} - 1} \left[ \ln \left( \frac{b_0}{\omega} \right) - 1 \right],
\end{equation}
where we used the identity $|\Gamma(1+ix)|^2 = \pi x/\sinh(\pi x)$. Here $b_0$ is associated to a finite frequency introduced as an upper-limit cut-off for the integral. We notice that the resulting expression for $N_\omega$ has the same structure as the one found in the commutative case. However, as mentioned before, the noncommutative effects are encoded in the surface gravity $\kappa_+$; explicitly, we have for the noncommutative-inspired RN black hole~\cite{Nicolini:2008aj}
\begin{equation}
2\kappa_+ = \frac 1{r_+} \left[ 1 - \frac {r_+^3 e^{-r_+^2/4\theta}}{4\theta^{3/2} \gamma \left( \dfrac 32, \dfrac {r_+^2}{4\theta} \right)} \right] - \frac {4q^2}{\pi r_+^3} \left[ \gamma^2 \left( \dfrac 32, \dfrac {r_+^2}{4\theta} \right) + \frac {r_+^3 e^{-r_+^2/4\theta}}{16\theta^{3/2} \gamma \left( \dfrac 32, \dfrac {r_+^2}{4\theta} \right)} F(r_+) \right],
\end{equation}
with $F(r_+)= \gamma^2(1/2,r_+^2/4\theta)-r_+ \gamma(1/2,r_+^2/2\theta)/\sqrt{2\theta}$.
We show in Fig.~\ref{partprod} the behaviour of the emitted particle spectrum for a fixed value of the noncommutative parameter $\theta$. We notice that the degeneracy of the spectrum in the classical RN black hole is removed, with the spectrum associated with different values of the charge clearly identified. Furthermore, in the noncommutative RN black hole, particle production is strongly suppressed for large frequencies, even when a small charge is present; we also infer that for a fixed value of the charge $q$, this effect is less noticeable as $\theta$ decreases and we approach the commutative limit. 

\section{Conclusions}

In this work, we applied the MM model to a noncommutative-inspired RN black hole to compute its Bogoliubov emission coefficient. For this purpose, a modified truncated trajectory was determined, which gives the essential physics at late times \cite{Good:2020qsy}, allowing us to analyse the number of particles as a function of the noncommutative parameter, which in turn led us to obtain a characteristic particle emission spectrum. We obtain thus a Planck distribution for the noncommutative-inspired RN black hole, similar to the result for the classical spectrum, but with the surface gravity modified by noncommutative contributions.

The main result of this work shows that, when comparing the classical case with the noncommutative scenario, noncommutativity affects particle production by reducing the number of particles emitted from external horizons as the noncommutative parameter increases, especially for large frequencies, thereby emulating the effect of the electric charge in the classical spectrum but more strongly. In fact, for a fixed value of the noncommutative parameter, there is a clear separation among the spectra for different charges, contrary to the situation in the classical case. This may consequently indicate why noncommutative-inspired black holes become colder than their classical counterparts.

It is worth noticing that the noncommutative spectrum reproducing the qualitative behaviour of electrically charged black holes can also be associated with rotating classical solutions \cite{Foo_2021}. This then reveals an interesting relation among classical thermodynamic variables $M, Q, a$ and the noncommutative parameter $\theta$ in the context of Hawking radiation, which is understandable, since these quantities are thermodynamic variables typically used for determining fundamental equations \cite{Misner:1973prb, Maya:2026gaz}. Although such thermodynamic variables follow quasihomogeneous relations that allow them to be identified \cite{Maya:2026gaz}, they cannot be easily distinguished through particle production analysis since they play an analogous qualitative role.

Since the noncommutative parameter exhibits the behaviour of an extensive quantity, it is reasonable to expect that noncommutative effects extend the expected duration of the evaporation process~\cite{Nicolini:2008aj}. However, to pursue this idea within a full thermodynamic framework, we should bear in mind that smeared Gaussian distributions of mass modify thermodynamic laws in a non-trivial and cumbersome way, so that a perturbative treatment is sometimes preferred~\cite{Banerjee:2011ljy}. This points out to considering alternative noncommutative distributions of mass and charge, such as Lorentzian ones, to facilitate calculations and gain further insight into these issues; see, for example~\cite{Filho:2024zxx}, where the analysis of the emission spectrum of a Schwarzschild-like black hole sourced by this kind of distribution is in agreement with our results.

\acknowledgments
A. Maya would like to thank SECIHTI for the postgraduate scholarship support (CVU No. 1031873).

\bibliography{refart}

\begin{thebibliography}{39}
\expandafter\ifx\csname natexlab\endcsname\relax\def\natexlab#1{#1}\fi
\expandafter\ifx\csname bibnamefont\endcsname\relax
  \def\bibnamefont#1{#1}\fi
\expandafter\ifx\csname bibfnamefont\endcsname\relax
  \def\bibfnamefont#1{#1}\fi
\expandafter\ifx\csname citenamefont\endcsname\relax
  \def\citenamefont#1{#1}\fi
\expandafter\ifx\csname url\endcsname\relax
  \def\url#1{\texttt{#1}}\fi
\expandafter\ifx\csname urlprefix\endcsname\relax\def\urlprefix{URL }\fi
\providecommand{\bibinfo}[2]{#2}
\providecommand{\eprint}[2][]{\url{#2}}

\bibitem[{\citenamefont{Bardeen et~al.}(1973)\citenamefont{Bardeen, Carter, and
  Hawking}}]{Bardeen:1973gs}
\bibinfo{author}{\bibfnamefont{J.~M.} \bibnamefont{Bardeen}},
  \bibinfo{author}{\bibfnamefont{B.}~\bibnamefont{Carter}}, \bibnamefont{and}
  \bibinfo{author}{\bibfnamefont{S.~W.} \bibnamefont{Hawking}},
  \bibinfo{journal}{Commun. Math. Phys.} \textbf{\bibinfo{volume}{31}},
  \bibinfo{pages}{161} (\bibinfo{year}{1973}).

\bibitem[{\citenamefont{Hawking}(1975)}]{Hawking:1975vcx}
\bibinfo{author}{\bibfnamefont{S.~W.} \bibnamefont{Hawking}},
  \bibinfo{journal}{Commun. Math. Phys.} \textbf{\bibinfo{volume}{43}},
  \bibinfo{pages}{199} (\bibinfo{year}{1975}), \bibinfo{note}{[Erratum:
  Commun.Math.Phys. 46, 206 (1976)]}.

\bibitem[{\citenamefont{Davies}(1975)}]{Davies:1974th}
\bibinfo{author}{\bibfnamefont{P.~C.~W.} \bibnamefont{Davies}},
  \bibinfo{journal}{J. Phys. A} \textbf{\bibinfo{volume}{8}},
  \bibinfo{pages}{609} (\bibinfo{year}{1975}).

\bibitem[{\citenamefont{Wald}(1975)}]{Wald:1975kc}
\bibinfo{author}{\bibfnamefont{R.~M.} \bibnamefont{Wald}},
  \bibinfo{journal}{Commun. Math. Phys.} \textbf{\bibinfo{volume}{45}},
  \bibinfo{pages}{9} (\bibinfo{year}{1975}).

\bibitem[{\citenamefont{Unruh}(1976)}]{Unruh:1976db}
\bibinfo{author}{\bibfnamefont{W.~G.} \bibnamefont{Unruh}},
  \bibinfo{journal}{Phys. Rev. D} \textbf{\bibinfo{volume}{14}},
  \bibinfo{pages}{870} (\bibinfo{year}{1976}).

\bibitem[{\citenamefont{Page}(1993)}]{Page:1993wv}
\bibinfo{author}{\bibfnamefont{D.~N.} \bibnamefont{Page}},
  \bibinfo{journal}{Phys. Rev. Lett.} \textbf{\bibinfo{volume}{71}},
  \bibinfo{pages}{3743} (\bibinfo{year}{1993}), \eprint{hep-th/9306083}.

\bibitem[{\citenamefont{Ryu and Takayanagi}(2006{\natexlab{a}})}]{Ryu:2006bv}
\bibinfo{author}{\bibfnamefont{S.}~\bibnamefont{Ryu}} \bibnamefont{and}
  \bibinfo{author}{\bibfnamefont{T.}~\bibnamefont{Takayanagi}},
  \bibinfo{journal}{Phys. Rev. Lett.} \textbf{\bibinfo{volume}{96}},
  \bibinfo{pages}{181602} (\bibinfo{year}{2006}{\natexlab{a}}),
  \eprint{hep-th/0603001}.

\bibitem[{\citenamefont{Ryu and Takayanagi}(2006{\natexlab{b}})}]{Ryu:2006ef}
\bibinfo{author}{\bibfnamefont{S.}~\bibnamefont{Ryu}} \bibnamefont{and}
  \bibinfo{author}{\bibfnamefont{T.}~\bibnamefont{Takayanagi}},
  \bibinfo{journal}{JHEP} \textbf{\bibinfo{volume}{08}}, \bibinfo{pages}{045}
  (\bibinfo{year}{2006}{\natexlab{b}}), \eprint{hep-th/0605073}.

\bibitem[{\citenamefont{Page}(2013)}]{Page:2013dx}
\bibinfo{author}{\bibfnamefont{D.~N.} \bibnamefont{Page}},
  \bibinfo{journal}{JCAP} \textbf{\bibinfo{volume}{09}}, \bibinfo{pages}{028}
  (\bibinfo{year}{2013}), \eprint{1301.4995}.

\bibitem[{\citenamefont{Almheiri
  et~al.}(2020{\natexlab{a}})\citenamefont{Almheiri, Hartman, Maldacena,
  Shaghoulian, and Tajdini}}]{Almheiri:2019qdq}
\bibinfo{author}{\bibfnamefont{A.}~\bibnamefont{Almheiri}},
  \bibinfo{author}{\bibfnamefont{T.}~\bibnamefont{Hartman}},
  \bibinfo{author}{\bibfnamefont{J.}~\bibnamefont{Maldacena}},
  \bibinfo{author}{\bibfnamefont{E.}~\bibnamefont{Shaghoulian}},
  \bibnamefont{and} \bibinfo{author}{\bibfnamefont{A.}~\bibnamefont{Tajdini}},
  \bibinfo{journal}{JHEP} \textbf{\bibinfo{volume}{05}}, \bibinfo{pages}{013}
  (\bibinfo{year}{2020}{\natexlab{a}}), \eprint{1911.12333}.

\bibitem[{\citenamefont{Almheiri
  et~al.}(2020{\natexlab{b}})\citenamefont{Almheiri, Mahajan, Maldacena, and
  Zhao}}]{Almheiri:2019hni}
\bibinfo{author}{\bibfnamefont{A.}~\bibnamefont{Almheiri}},
  \bibinfo{author}{\bibfnamefont{R.}~\bibnamefont{Mahajan}},
  \bibinfo{author}{\bibfnamefont{J.}~\bibnamefont{Maldacena}},
  \bibnamefont{and} \bibinfo{author}{\bibfnamefont{Y.}~\bibnamefont{Zhao}},
  \bibinfo{journal}{JHEP} \textbf{\bibinfo{volume}{03}}, \bibinfo{pages}{149}
  (\bibinfo{year}{2020}{\natexlab{b}}), \eprint{1908.10996}.

\bibitem[{\citenamefont{Penington et~al.}(2022)\citenamefont{Penington,
  Shenker, Stanford, and Yang}}]{Penington:2019kki}
\bibinfo{author}{\bibfnamefont{G.}~\bibnamefont{Penington}},
  \bibinfo{author}{\bibfnamefont{S.~H.} \bibnamefont{Shenker}},
  \bibinfo{author}{\bibfnamefont{D.}~\bibnamefont{Stanford}}, \bibnamefont{and}
  \bibinfo{author}{\bibfnamefont{Z.}~\bibnamefont{Yang}},
  \bibinfo{journal}{JHEP} \textbf{\bibinfo{volume}{03}}, \bibinfo{pages}{205}
  (\bibinfo{year}{2022}), \eprint{1911.11977}.

\bibitem[{\citenamefont{Maldacena}(2020)}]{Maldacena:2020ady}
\bibinfo{author}{\bibfnamefont{J.}~\bibnamefont{Maldacena}},
  \bibinfo{journal}{Nature Rev. Phys.} \textbf{\bibinfo{volume}{2}},
  \bibinfo{pages}{123} (\bibinfo{year}{2020}).

\bibitem[{\citenamefont{Spallucci and Smailagic}(2014)}]{Spallucci:2014kua}
\bibinfo{author}{\bibfnamefont{E.}~\bibnamefont{Spallucci}} \bibnamefont{and}
  \bibinfo{author}{\bibfnamefont{A.}~\bibnamefont{Smailagic}}, in
  \emph{\bibinfo{booktitle}{Advances in Black Holes Research}}, edited by
  \bibinfo{editor}{\bibfnamefont{A.}~\bibnamefont{Barton}}
  (\bibinfo{publisher}{Nova Science Publishers, NY, USA},
  \bibinfo{year}{2014}), \eprint{1410.1706}.

\bibitem[{\citenamefont{Nicolini et~al.}(2006)\citenamefont{Nicolini,
  Smailagic, and Spallucci}}]{Nicolini:2005vd}
\bibinfo{author}{\bibfnamefont{P.}~\bibnamefont{Nicolini}},
  \bibinfo{author}{\bibfnamefont{A.}~\bibnamefont{Smailagic}},
  \bibnamefont{and}
  \bibinfo{author}{\bibfnamefont{E.}~\bibnamefont{Spallucci}},
  \bibinfo{journal}{Phys. Lett.} \textbf{\bibinfo{volume}{B632}},
  \bibinfo{pages}{547} (\bibinfo{year}{2006}), \eprint{gr-qc/0510112}.

\bibitem[{\citenamefont{Nicolini}(2005)}]{Nicolini:2005gy}
\bibinfo{author}{\bibfnamefont{P.}~\bibnamefont{Nicolini}},
  \bibinfo{journal}{Bled Workshops Phys.} \textbf{\bibinfo{volume}{6}},
  \bibinfo{pages}{79} (\bibinfo{year}{2005}), \eprint{hep-th/0510203}.

\bibitem[{\citenamefont{Banerjee et~al.}(2010)\citenamefont{Banerjee,
  Gangopadhyay, and Modak}}]{Banerjee:2009xx}
\bibinfo{author}{\bibfnamefont{R.}~\bibnamefont{Banerjee}},
  \bibinfo{author}{\bibfnamefont{S.}~\bibnamefont{Gangopadhyay}},
  \bibnamefont{and} \bibinfo{author}{\bibfnamefont{S.~K.} \bibnamefont{Modak}},
  \bibinfo{journal}{Phys. Lett.} \textbf{\bibinfo{volume}{B686}},
  \bibinfo{pages}{181} (\bibinfo{year}{2010}), \eprint{0911.2123 [hep-th]}.

\bibitem[{\citenamefont{Nicolini}(2009)}]{Nicolini:2008aj}
\bibinfo{author}{\bibfnamefont{P.}~\bibnamefont{Nicolini}},
  \bibinfo{journal}{Int. J. Mod. Phys.} \textbf{\bibinfo{volume}{A24}},
  \bibinfo{pages}{1229} (\bibinfo{year}{2009}), \eprint{0807.1939 [hep-th]}.

\bibitem[{\citenamefont{Nozari and Fazlpour}(2008)}]{Nozari:2006rt}
\bibinfo{author}{\bibfnamefont{K.}~\bibnamefont{Nozari}} \bibnamefont{and}
  \bibinfo{author}{\bibfnamefont{B.}~\bibnamefont{Fazlpour}},
  \bibinfo{journal}{Acta Phys. Polon. B} \textbf{\bibinfo{volume}{39}},
  \bibinfo{pages}{1363} (\bibinfo{year}{2008}), \eprint{gr-qc/0608077}.

\bibitem[{\citenamefont{Maya and Quevedo}(2026)}]{Maya:2026gaz}
\bibinfo{author}{\bibfnamefont{A.}~\bibnamefont{Maya}} \bibnamefont{and}
  \bibinfo{author}{\bibfnamefont{H.}~\bibnamefont{Quevedo}},
  \bibinfo{journal}{Int. J. Mod. Phys. D} \textbf{\bibinfo{volume}{35}},
  \bibinfo{pages}{2650047} (\bibinfo{year}{2026}), \eprint{2601.17081}.

\bibitem[{\citenamefont{Smailagic and Spallucci}(2003)}]{Smailagic:2003rp}
\bibinfo{author}{\bibfnamefont{A.}~\bibnamefont{Smailagic}} \bibnamefont{and}
  \bibinfo{author}{\bibfnamefont{E.}~\bibnamefont{Spallucci}},
  \bibinfo{journal}{J.Phys.} \textbf{\bibinfo{volume}{A36}},
  \bibinfo{pages}{L517} (\bibinfo{year}{2003}), \eprint{hep-th/0308193}.

\bibitem[{\citenamefont{Smailagic and Spallucci}(2004)}]{Smailagic:2004yy}
\bibinfo{author}{\bibfnamefont{A.}~\bibnamefont{Smailagic}} \bibnamefont{and}
  \bibinfo{author}{\bibfnamefont{E.}~\bibnamefont{Spallucci}},
  \bibinfo{journal}{J.Phys.} \textbf{\bibinfo{volume}{A37}}, \bibinfo{pages}{1}
  (\bibinfo{year}{2004}), \eprint{hep-th/0406174}.

\bibitem[{\citenamefont{Modesto and Nicolini}(2010)}]{Modesto:2010rv}
\bibinfo{author}{\bibfnamefont{L.}~\bibnamefont{Modesto}} \bibnamefont{and}
  \bibinfo{author}{\bibfnamefont{P.}~\bibnamefont{Nicolini}},
  \bibinfo{journal}{Phys. Rev.} \textbf{\bibinfo{volume}{D82}},
  \bibinfo{pages}{104035} (\bibinfo{year}{2010}), \eprint{1005.5605 [gr-qc]}.

\bibitem[{\citenamefont{Parikh and Wilczek}(2000)}]{Parikh:1999mf}
\bibinfo{author}{\bibfnamefont{M.~K.} \bibnamefont{Parikh}} \bibnamefont{and}
  \bibinfo{author}{\bibfnamefont{F.}~\bibnamefont{Wilczek}},
  \bibinfo{journal}{Phys. Rev. Lett.} \textbf{\bibinfo{volume}{85}},
  \bibinfo{pages}{5042} (\bibinfo{year}{2000}), \eprint{hep-th/9907001}.

\bibitem[{\citenamefont{Chakraborty and Saha}(2014)}]{Chakraborty:2014gwx}
\bibinfo{author}{\bibfnamefont{S.}~\bibnamefont{Chakraborty}} \bibnamefont{and}
  \bibinfo{author}{\bibfnamefont{S.}~\bibnamefont{Saha}},
  \bibinfo{journal}{Adv. High Energy Phys.} \textbf{\bibinfo{volume}{2014}},
  \bibinfo{pages}{168487} (\bibinfo{year}{2014}), \eprint{1702.06991}.

\bibitem[{\citenamefont{Moore}(1970)}]{Moore:1970tmc}
\bibinfo{author}{\bibfnamefont{G.~T.} \bibnamefont{Moore}},
  \bibinfo{journal}{J. Math. Phys.} \textbf{\bibinfo{volume}{11}},
  \bibinfo{pages}{2679} (\bibinfo{year}{1970}).

\bibitem[{\citenamefont{DeWitt}(1975)}]{DeWitt:1975ys}
\bibinfo{author}{\bibfnamefont{B.~S.} \bibnamefont{DeWitt}},
  \bibinfo{journal}{Phys. Rept.} \textbf{\bibinfo{volume}{19}},
  \bibinfo{pages}{295} (\bibinfo{year}{1975}).

\bibitem[{\citenamefont{Davies and Fulling}(1976)}]{Davies:1976hi}
\bibinfo{author}{\bibfnamefont{P.~C.~W.} \bibnamefont{Davies}}
  \bibnamefont{and} \bibinfo{author}{\bibfnamefont{S.~A.}
  \bibnamefont{Fulling}}, \bibinfo{journal}{Proc. Roy. Soc. Lond. A}
  \textbf{\bibinfo{volume}{348}}, \bibinfo{pages}{393} (\bibinfo{year}{1976}).

\bibitem[{\citenamefont{Davies and Fulling}(1977)}]{Davies:1977yv}
\bibinfo{author}{\bibfnamefont{P.~C.~W.} \bibnamefont{Davies}}
  \bibnamefont{and} \bibinfo{author}{\bibfnamefont{S.~A.}
  \bibnamefont{Fulling}}, \bibinfo{journal}{Proc. Roy. Soc. Lond. A}
  \textbf{\bibinfo{volume}{356}}, \bibinfo{pages}{237} (\bibinfo{year}{1977}).

\bibitem[{\citenamefont{Carlitz and Willey}(1987)}]{Carlitz:1986nh}
\bibinfo{author}{\bibfnamefont{R.~D.} \bibnamefont{Carlitz}} \bibnamefont{and}
  \bibinfo{author}{\bibfnamefont{R.~S.} \bibnamefont{Willey}},
  \bibinfo{journal}{Phys. Rev. D} \textbf{\bibinfo{volume}{36}},
  \bibinfo{pages}{2327} (\bibinfo{year}{1987}).

\bibitem[{\citenamefont{Wilczek}(1993)}]{Wilczek:1993jn}
\bibinfo{author}{\bibfnamefont{F.}~\bibnamefont{Wilczek}}, in
  \emph{\bibinfo{booktitle}{{International Symposium on Black holes, Membranes,
  Wormholes and Superstrings}}} (\bibinfo{year}{1993}),
  \eprint{hep-th/9302096}.

\bibitem[{\citenamefont{Good et~al.}(2013)\citenamefont{Good, Anderson, and
  Evans}}]{Good:2013lca}
\bibinfo{author}{\bibfnamefont{M.~R.~R.} \bibnamefont{Good}},
  \bibinfo{author}{\bibfnamefont{P.~R.} \bibnamefont{Anderson}},
  \bibnamefont{and} \bibinfo{author}{\bibfnamefont{C.~R.} \bibnamefont{Evans}},
  \bibinfo{journal}{Phys. Rev. D} \textbf{\bibinfo{volume}{88}},
  \bibinfo{pages}{025023} (\bibinfo{year}{2013}), \eprint{1303.6756}.

\bibitem[{\citenamefont{Good et~al.}(2017{\natexlab{a}})\citenamefont{Good,
  Anderson, and Evans}}]{Good:2015jwa}
\bibinfo{author}{\bibfnamefont{M.~R.~R.} \bibnamefont{Good}},
  \bibinfo{author}{\bibfnamefont{P.~R.} \bibnamefont{Anderson}},
  \bibnamefont{and} \bibinfo{author}{\bibfnamefont{C.~R.} \bibnamefont{Evans}},
  in \emph{\bibinfo{booktitle}{{14th Marcel Grossmann Meeting on Recent
  Developments in Theoretical and Experimental General Relativity,
  Astrophysics, and Relativistic Field Theories}}}
  (\bibinfo{year}{2017}{\natexlab{a}}), vol.~\bibinfo{volume}{2}, pp.
  \bibinfo{pages}{1705--1708}, \eprint{1507.05048}.

\bibitem[{\citenamefont{Good and Ong}(2020)}]{Good:2020qsy}
\bibinfo{author}{\bibfnamefont{M.~R.~R.} \bibnamefont{Good}} \bibnamefont{and}
  \bibinfo{author}{\bibfnamefont{Y.~C.} \bibnamefont{Ong}},
  \bibinfo{journal}{Eur. Phys. J. C} \textbf{\bibinfo{volume}{80}},
  \bibinfo{pages}{1169} (\bibinfo{year}{2020}), \eprint{2004.03916}.

\bibitem[{\citenamefont{Good et~al.}(2017{\natexlab{b}})\citenamefont{Good,
  Yelshibekov, and Ong}}]{Good:2016atu}
\bibinfo{author}{\bibfnamefont{M.~R.~R.} \bibnamefont{Good}},
  \bibinfo{author}{\bibfnamefont{K.}~\bibnamefont{Yelshibekov}},
  \bibnamefont{and} \bibinfo{author}{\bibfnamefont{Y.~C.} \bibnamefont{Ong}},
  \bibinfo{journal}{JHEP} \textbf{\bibinfo{volume}{3}}, \bibinfo{pages}{13}
  (\bibinfo{year}{2017}{\natexlab{b}}), \eprint{1611.00809}.

\bibitem[{\citenamefont{Foo and Good}(2021)}]{Foo_2021}
\bibinfo{author}{\bibfnamefont{J.}~\bibnamefont{Foo}} \bibnamefont{and}
  \bibinfo{author}{\bibfnamefont{M.~R.} \bibnamefont{Good}},
  \bibinfo{journal}{Journal of Cosmology and Astroparticle Physics}
  \textbf{\bibinfo{volume}{2021}}, \bibinfo{pages}{019–019}
  (\bibinfo{year}{2021}), ISSN \bibinfo{issn}{1475-7516},
  \urlprefix\url{http://dx.doi.org/10.1088/1475-7516/2021/01/019}.

\bibitem[{\citenamefont{Misner et~al.}(1973)\citenamefont{Misner, Thorne, and
  Wheeler}}]{Misner:1973prb}
\bibinfo{author}{\bibfnamefont{C.~W.} \bibnamefont{Misner}},
  \bibinfo{author}{\bibfnamefont{K.~S.} \bibnamefont{Thorne}},
  \bibnamefont{and} \bibinfo{author}{\bibfnamefont{J.~A.}
  \bibnamefont{Wheeler}}, \emph{\bibinfo{title}{{Gravitation}}}
  (\bibinfo{publisher}{W. H. Freeman}, \bibinfo{address}{San Francisco},
  \bibinfo{year}{1973}), ISBN \bibinfo{isbn}{978-0-7167-0344-0,
  978-0-691-17779-3}.

\bibitem[{\citenamefont{Banerjee and Gangopadhyay}(2011)}]{Banerjee:2011ljy}
\bibinfo{author}{\bibfnamefont{R.}~\bibnamefont{Banerjee}} \bibnamefont{and}
  \bibinfo{author}{\bibfnamefont{S.}~\bibnamefont{Gangopadhyay}},
  \bibinfo{journal}{Gen. Rel. Grav.} \textbf{\bibinfo{volume}{43}},
  \bibinfo{pages}{3201} (\bibinfo{year}{2011}), \eprint{1008.1683}.

\bibitem[{\citenamefont{Filho et~al.}(2024)\citenamefont{Filho, Nascimento,
  Petrov, Porf{\'\i}rio, and {\"O}vg{\"u}n}}]{Filho:2024zxx}
\bibinfo{author}{\bibfnamefont{A.~A.~A.} \bibnamefont{Filho}},
  \bibinfo{author}{\bibfnamefont{J.~R.} \bibnamefont{Nascimento}},
  \bibinfo{author}{\bibfnamefont{A.~Y.} \bibnamefont{Petrov}},
  \bibinfo{author}{\bibfnamefont{P.~J.} \bibnamefont{Porf{\'\i}rio}},
  \bibnamefont{and}
  \bibinfo{author}{\bibfnamefont{A.}~\bibnamefont{{\"O}vg{\"u}n}},
  \bibinfo{journal}{Phys. Dark Univ.} \textbf{\bibinfo{volume}{46}},
  \bibinfo{pages}{101630} (\bibinfo{year}{2024}), \eprint{2406.12015}.

\end{thebibliography}

\end{document}